\documentstyle[sprocl]{article}
\def\Journal#1#2#3#4{{#1} {\bf #2}, #3 (#4)}
\def\ApJ{\em ApJ}

\def\NPB{{\em Nucl. Phys.} B}

\def\PRD{{\em Phys. Rev.} D}
\def\al{\alpha}
\def\be{\begin{equation}}
\def\ee{\end{equation}}
\begin{document}
\title{GENERATION OF GRAVITATIONAL WAVES AND SCALAR PERTURBATIONS\\ IN
INFLATION WITH EFFECTIVE $\Lambda$-TERM \\AND T/S STORY}
\author{ V.N.LUKASH and E.V.MIKHEEVA }
\address{Astro Space Center of P.N.Lebedev Physical Institute,\\
Profsoyuznaya Street 84/32, Moscow, 117810, Russia}
\maketitle
\abstracts{ We argue that gravitational wave contribution
to the cosmic microwave background anisotropy at angular scale
$\sim 10^0$ may exceed 50 \% for some models of hybrid inflation
producing standard cosmology with the density perturbation slope $n\simeq 1$.}

\section{Introduction}
Two types of metric perturbations $h_{\al\beta}$ generating at inflation
driven by scalar field, can contribute to 
large-scale cosmic microwave background anisotropy through the SW 
effect\cite{sw}. They are the Scalar (density) and 
Tensor (gravitational waves) perturbations which are presented as follows:
$h_{\al\beta}/2=A\delta_{\al\beta}+B_{,\al\beta}+G_{\al\beta}$.
The first and second terms correspond to the scalar mode whereas the
third one is for the tensor one ($G^{\al}_{\al}=G^{\beta}_{\al,\beta}=0$).
Their contribution into $\Delta T/T$ may be standardly separated
over $S$ and $T$ parts:
\be 
\Big\langle
\left( 
\frac{\Delta T}{T} 
\right)^2_{10^0}
\Big\rangle
=
\sum_{\ell} 
\langle 
a_{\ell}^2 
\rangle
=
\sum_{\ell} 
\left( 
\langle 
a_{\ell}^2 
\rangle{}_S
+
\langle 
a_{\ell}^2
\rangle{}_T
\right)
=
S
+
T. 
\ee

The available experimental data are not sufficient
yet to recognize and find gravitational wave fraction in $\Delta T/T$
(that could be possible, in principle, by means of a joint
analysis of the dependence of the temperature fluctuation
spectrum on scale, the amplitude of Dopper peak, etc., which is
the matter of future observations). Nevertheless, today we can
evaluate $T/S$ theoretically for various inflationary models to
test the possible effect and see how large it may deviate from
model to model and which particular properties of the model
correlate with large $T/S$. Actually, we construct a
counterexample to the common prejudice that ``$T/S$ {\em is negligible
for the Harrison-Zel'dovich spectrum}'', introducing here a
general class of models which face the opposite conclusion:
``$T/S$ {\em may be about or even larger than 1 for} $n\simeq
1$''.

\section{ Generation of cosmological perturbations in inflationary
model with effective $\Lambda$-term}
A lot of inflation models has been discussed in the literature\cite{rocky} 
for $T/S$. The result was that large $T/S$ could be achieved at the
expense of the rejection from the Harrison-Zeldovich spectrum: 
$T/S\ge 1$ for $n_S\le 0.8$, the ``red'' spectra. Thus, it would be 
interesting to study possible $T/S$ in models producing ``blue'' ($n_S>1$)
spectra of density perturbations. Here we present a general class of 
models based on the only scalar field with the potential:
\be
V=V_0+\frac{m^2\varphi^2}{2},
\ee
where $V_0$ and $m$ are constants, $V_0$ describe the effective (metastable) 
$\Lambda$-term. The mechanism of its decay is not fixed here and may be 
arbitrary, for example, with help of another scalar field like in the 
hybrid 
inflation\cite{linde}.

It is clear that the inflaton dynamics in the potential (2) has two regims 
and, consequently, the spectra of cosmological perturbations has two 
asymptotics. The regims are separated by the critical value of field 
$\varphi_{cr}$ at which the first term is equal to the second 
$\varphi^2_{cr}=2V_0/m^2$.

At the first stage (we call it ``red'' asymptotic --- by the form of the
density perturbation spectrum) $\varphi$ is large and the massive term 
dominantes the evolution. This is similar to the case of well-known chaotic 
inflation, so it is easy to write down the spectra of the perturbations\cite{aw,luk} 
as follows ($c=\hbar=8\pi G=1$):
\be
q^r_k=\frac{m\varphi^2}{4\sqrt{6}\pi},
\;\;\;\;
h^r_k=\frac{m\varphi}{2\sqrt{3}\pi},
\ee

For $\varphi<\varphi_{cr}$ the constant term predominates. Here we obtain 
the ``blue'' asymptotic spectrum of density perturbations:
\be
q^b_k=\frac{H_0c_n^2\varphi^2_{cr}}{4\pi\varphi},
\;\;\;\;
h^b_k=\frac{H_0}{\sqrt{2}\pi},
\ee
where $H_0=(V_0/3)^{1/2}$, 
$c_n^2=2^{1/2-n_S/2}(7-n_S)\Gamma(2-n_S/2)(3\sqrt{\pi})^{-1}$, 
$\Gamma(x)$ is gamma function, $q$ coincides with $A$ in comoving frame
and $h^2_k$ is the spectrum of $G_{\al\beta}G^{\al\beta}$. $\varphi$  in
(3), (4) is taken at horizon crossing $k=a(\varphi)H(\varphi)$, $a$ and 
$H=(V/3)^{1/2}$ are scale and Hubble factors, respectively. 
Note that we have the following relation 
between the spectrum slope at the ``blue'' asymptotic $n_S$ and 
$\varphi_{cr}$: $(n_S-1)(7-n_S)=24\varphi^{-2}_{cr}$, $\varphi_{cr}^2\geq 
8/3$, $1<n_S<4$. 

As for gravitational waves their spectrum remains universal for any $k$, but
the density perturbation spectrum requires approximation (fitting both 
asymptotics 
and becoming exact in the slow-roll limit):
\be
q_k=\frac{H_0\varphi_{cr}}{4\pi y}(1+y^2)^{1/2}(c_n^2+y^2),
\;\;\;\;
h_k=\frac{H_0}{\sqrt 2\pi}(1+y^2)^{1/2},
\ee
where $y=\varphi/\varphi_{cr}$.
It is easy to see that the ratio of tensor to scalar spectra achieves 
its maximum at $\varphi\simeq\varphi_{cr}$, where the minimum
of density perturbation power occurs:
\be
\frac{h_k}{q_k}=\frac{2\sqrt{2}}{\varphi_{cr}}\frac{y}{c_n^2+y^2}\leq
\frac{\sqrt{2}}{c_n\varphi_{cr}}=D_n,
\ee
where $D_2=0.8$, $D_3=1$. We may suppose that here $T/S$ gets its maximum as 
well.

$T/S$ was calculated for potential (2) with DMR COBE window function. We 
obtained that $T/S$ function is a two-parametric one and depends on $n_S$ and
$k_{cr}$, the latter corresponds to $\varphi_{cr}$. However, the earlier
introduced approximation formula\cite{rocky}, $T/S\simeq -6n_T$, holds in
our case as well if both $T/S$ and local $|n_T|$ are taken at their maxima. 
We found a phenomenological relation between
$n_{T}$ and $n_S$: $|n_{T}|\simeq (n_S-1)/4$, thus, another formula for $T/S$ can be
proposed for this model: $T/S|_{max}\simeq 1.5(n_S-1)$.
\section{Conclusions}
Finally, we conclude: 
The inflation model predicting $T/S>1$ has been constructed. 
A property of this model is ``blue'' spectrum of density perturbations 
for scales $k \gg k_{cr}$. However in the region, where $T/S\ge 1$, 
$k\sim k_{cr}$ and therefore the locally observed spectrum of cosmological 
perturbations is close to scale-invariant ($n\simeq 1$).
%
%
%\section*{Acknowledgements}
\vspace*{3mm}

The work was supported partly by the Russian Foundation for
Fundamental Research (project code 96-02-16689-a) and COSMION
(cosmomicrophysics).


\section*{References}
\begin{thebibliography}{99}
\bibitem{sw} R.K. Sachs and A.M. Wolfe,\Journal{\ApJ}{147}{73}{1967}.
\bibitem{rocky} J.E.Lidsey {\it et al.}, astro-ph/9508078.
\bibitem{linde} A. Linde, \Journal{\PRD}{49}{748}{1994}.
\bibitem{aw} L.F. Abbott and M.B. Wise, \Journal{\NPB}{244}{279}{1984}.
\bibitem{luk} V.N. Lukash, in {\em Cosmology and Gravitation II}, ed. M.Novello (Editions Frontieres, 1996).
\end{thebibliography}
\end{document}